\documentclass[12pt]{iopart}
\usepackage{graphicx}
\usepackage{multirow}
\usepackage{epsfig}
\usepackage{color}

\usepackage{iopams,epsfig}

\newcommand{\bea}{\begin{eqnarray}}
\newcommand{\eea}{\end{eqnarray}}
\newcommand{\ba}{\begin{array}}
\newcommand{\ea}{\end{array}}

\newcommand{\be}{\begin{equation}}

\newcommand{\ee}{\end{equation}}
\newcommand{\bt}{\begin{teo}}
\newcommand{\et}{\end{teo}}

\begin{document}

\title[]
{Statistical properties of the energy in time-dependent homogeneous power law potentials}

\author{Dimitris Andresas and Marko Robnik}

\address{CAMTP - Center for Applied Mathematics and Theoretical Physics,
University of Maribor, Krekova 2, SI-2000 Maribor, Slovenia, European Union}


\eads{dimitraklos@hotmail.com, Robnik@uni-mb.si}

\begin{abstract}
We study  classical 1D Hamilton systems with homogeneous power law potential 
and their statistical behaviour, assuming the microcanonical distribution of the initial
conditions and describing its change under monotonically increasing time-dependent 
function $a(t)$ (prefactor of the potential). 
Using the nonlinear WKB-like method by Papamikos and 
Robnik {\em J. Phys. A: Math. Theor. {\bf 44} (2012) 315102} 
and following a previous work by Papamikos G and Robnik M 
{\em J. Phys. A: Math. Theor. {\bf 45} (2011) 015206} we 
specifically analyze the mean energy, the variance and the 
adiabatic invariant (action) of the systems for large time 
$t\rightarrow\infty$ and we show that the mean energy 
and variance increase as powers of $a(t)$, while the action oscillates and finally 
remains constant. By means of a number of detailed case studies we show that 
the theoretical prediction is excellent which demonstrates the usefulness of the 
method in such applications.

\end{abstract}

\pacs{05.00, 05.45.-a, 05.45.Ac}

\section{Introduction} \label{Intro}

Time-dependent Hamiltonian systems \cite{Arnold, Lochak, Zaslavsky, Ott}
are very interesting and important dynamical models, 
where many important questions about their dynamical behaviour can be studied. 
While the energy of the system is not conserved, the Liouville theorem 
of course still applies and thus the phase space volume is preserved by the
flow. One of the central questions is the evolution of the energy of 
certain ensemble of initial conditions for such systems. In particular in time-periodic 
systems we can find a very rich behaviour, from integrability to full
chaoticity (ergodicity) and also the scenario in between, 
namely the cases of a mixed phase space, even in one dimensional systems
\cite{Zaslavsky,Chirikov,PapRob2011,PapSowRob2012}.
We will consider the general family of classical 1D Hamiltonian systems 
with homogeneous power law potential and quadratic kinetic energy, as follows,
\bea\label{GH}
H(q, p, t) = \frac{p^{2}}{2} + a(t) V(q),
\eea
where $p$ is the momentum, $a(t)\,V(q)=a(t)\frac{q^{2m}}{2m}$ the potential
as a function of the coordinate $q$ and time $t$, $a(t)$ is a time-dependent function 
and $m$ is an integer $m=1,2,\ldots$. In a recent work Papamikos and 
Robnik \cite{PapRob2012} 
have developed the first-order nonlinear WKB-like method for such
homogeneous power law potentials as an approximation of the general solution, 
which can be used successfully to generalize a 
series of studies on the time-dependent linear oscillator by Robnik and Romanovski
\cite{RR2006a,RR2006b,RRS2006,KR2007,RR2008}, where the rigorous linear
WKB method (to all orders) has been employed \cite{RR2000}. Using these tools 
we shall analyze the statistical properties of the energy of systems (\ref{GH}).

We are particularly interested in 
the time evolution of a microcanonical ensemble of initial conditions. 
Namely, the most natural ensemble, and the most important
one, is the microcanonical ensemble of the initial conditions,
because if we have a large ensemble of identical systems with the same ("prepared") energy,
and we do not have any further information about them, the uniform distribution with respect
to the canonical angle ("the phases") is the most appropriate one.
Moreover, in the context of the statistical mechanics, also for such
low-dimensional systems, the energy has a special status as the
state variable (see the introductory discussion in reference \cite{ABR}). 

If the evolution is ideal adiabatic (i.e. infinitely slow), 
then the adiabatic invariant,
which is also the action of the system, or the area inside the 
contour of constant energy in the phase space (divided by $2\pi$), 
is conserved, and this is precisely the adiabatic theorem on one-dimensional Hamilton
systems \cite{Arnold}, provided we do not cross a separatrix 
during the adiabatic process. The energy is sharply distributed (Dirac delta function)
and its value is fully determined by the value of the adiabatic invariant.
For faster changes of the system parameter the adiabatic invariant 
is no longer conserved and the energy becomes distributed over a certain
interval. For the linear oscillator this turns out to be precisely the arcsine 
distribution for {\em any frequency} $\omega(t)$ as a function of time $t$,
where in this case we have $m=1$ and $a(t) =\omega(t)^2$ in (\ref{GH}). 
It has also been proven in \cite{RR2006a} and further discussed in
\cite{RR2006b,RRS2006,KR2007,RR2008} that the value of the adiabatic invariant
(action) always increases, except in ideal adiabatic (=infinitely slow)
process, where it is constant. 

What happens for higher power law potentials  when the 
potential, namely $a(t)$, changes in time?  For the single parametric kick
(instantaneous discontinuous jump of the parameter $a(t)$) it has been
shown in \cite{PapRob2011} and further developed (even for a large family
of potentials) in \cite{ABR}, that the action at the average final energy  
increases under very general
conditions, thus exhibiting the so-called {\em PR property}. If the variation 
of $a(t)$ is slow, but not ideally adiabatic, the action can decrease, due
to the nonlinearity. The evolution of the energy and its distribution has
been studied there only for large times up to $T=1/\epsilon$, where
$\epsilon$ is the adiabtic parameter, and $T$ is the time interval of
the variation of $a(t)$.

In this paper we study the case where the time dependent function $a(t)$
increases monotonically and unbounded, very much beyond $1/\epsilon$, 
for the systems (\ref{GH}).
We will use higher order symplectic numerical techniques 
\cite{M1995,McL2002,HLW2006,LR2004,SSC1994,SY1996,Y1990,Y1993}, 
namely the 8th order symplectic integrator, 
and on the theoretical side the WKB-like nonlinear method for homogeneous 
power law potentials \cite{PapRob2012} 
in order to describe the evolution of a microcanonical ensemble 
of initial conditions for large times.
Along with the numerical energy distribution as a function of time, we
 calculate in particular the corresponding mean energy and the variance and we show that these
results agree with the general theory, which we will present in the next section.
 
In the special case of time evolution of a microcanonical ensemble of such systems 
with $a(t)=1 +\epsilon t$, the dynamics exhibits a structure consisting of an
energy layer (see figure 1).
\begin{figure} 
\center 
    \includegraphics[width=15cm]{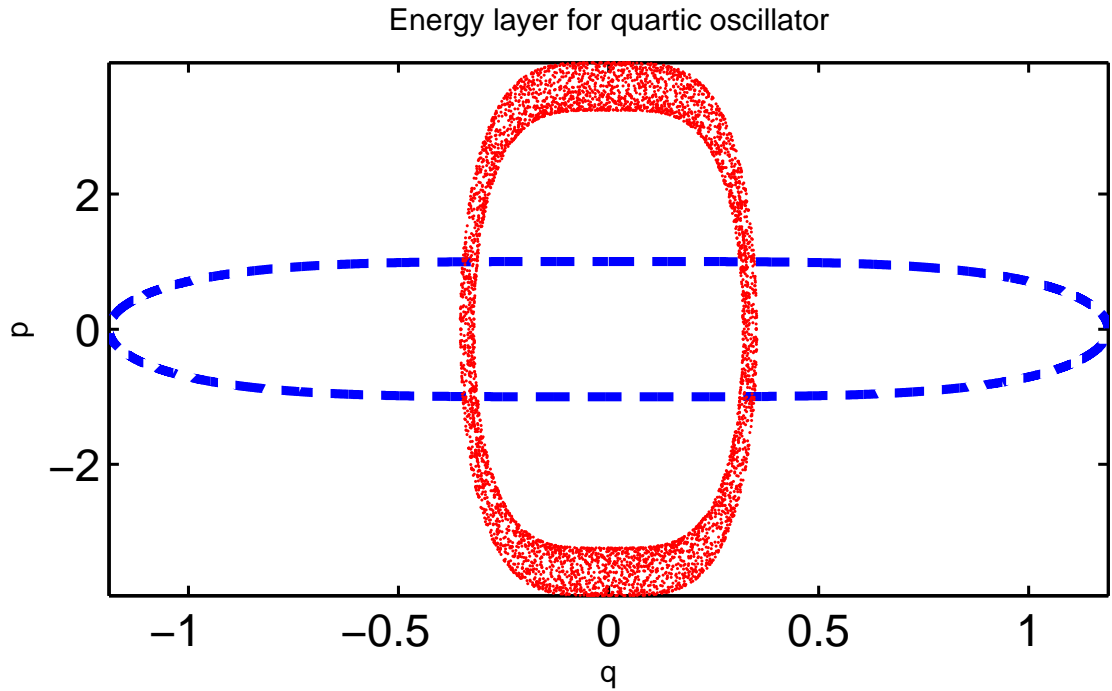}
    \caption{The energy layer for the quartic oscillator $(m=2)$ of 
    the initial microcacnonical ensemble of 5000
    points (dashed), with initial energy $E_0 = 0.5$ and $a(t) = 1 + \epsilon t$, 
    for $\epsilon = 1$ and $t=2000$. The layer evolves with time, becoming
monotonically thicker.}
\label{action2}
\end{figure}
We shall calculate the width of this layer as a function of
time using the nonlinear WKB method adapted for such
systems and we will explore the behaviour of the average energy and of the
variance of the energy for an initial microcanonical ensemble for large 
times $t\rightarrow\infty$. If $a(t)$ is a monotonic and unbounded 
increasing function of time, then both quantities are powers of $a(t)$.
In this limit the action  first oscillates and finally remains asymptotically 
constant, but 
assumes slightly larger value than its initial value. This is an interesting 
result for time-dependent Hamiltonian systems in a more general
context \cite{Rob2013,ABR}.
The distribution function of the final energy will be calculated numerically,
but cannot be derived theoretically, unlike the case of the linear oscillator $m=1$,
where Robnik, Romanovski and St\"ockmann have 
proven the explicit exact formula for the distribution, which is the
arcsine distribution \cite{RRS2006,RR2008}.

\section{General theory of time-dependent homogeneous power law potential} \label{General}

We consider the Hamilton systems with one degree of freedom in 
the form quadratic kinetic energy plus a homogeneous power law potential, as defined in equation ~\eref{GH}.
Using the nonlinear WKB-like method \cite{PapRob2012} to the first order approximation for such
nonlinear Hamiltonian systems, we have the following general solution,
\bea \label{GS}
q(t)&= a(t)^{-\frac{1}{2m+2}}g\left(\int^{t}_{t_0} a(x)^{\frac{1}{m+1}}dx\right), \nonumber \\
p(t)&= -\frac{a^{'}(t)\;(a(t))^{-\frac{1}{2m+2}-1}}{2m+2}g\left(\int^{t}_{t_0}a(x)^{\frac{1}{m+1}}dx\right) \nonumber \\
&+ (a(t))^{\frac{1}{2m+2}} g^{'}\left(\int^{t}_{t_0}a(x)^{\frac{1}{m+1}}dx\right),
\eea
where $g$ is the solution of the corresponding time-independent 
Hamiltonian system with $a(t)=1$ and $g^{'}$ is its first 
derivative with respect to time $t$. The function $g$ satisfies the following
differential equation,
\bea \label{ode}
g^{''} + f(g) = 0,
\eea
where, $f(g) = dV(g)/dg$.  
The differential equation \eref{ode} admits the first integral, namely the total energy of the oscillator, 
which is a function of the initial energy $E_0$ and the initial phase (canonical angle) $\phi_0$ of
the oscillator ~\eref{GH},
\bea
\quad \frac{(g^{'}(t))^{2}}{2} + \frac{(g(t))^{2m}}{2m} = c(\phi_{0}, E_0).
\eea
The dependence on $E_0$ and $\phi_0$ enters through the lower integration limit
in equation (\ref{GS}).
The expression for the energy of ~\eref{GH} is
\bea
E(t) &= a(t)^{\frac{1}{m+1}}\left[\frac{\left(g\left(\int^{t}_{t_{0}}a(x)^{\frac{1}{m+1}}dx\right)\right)^{2m}}{2m}+\frac{\left(g^{'}\left(\int^{t}_{t_{0}}a(x)^{\frac{1}{m+1}}dx\right)\right)^{2}}{2}\right] \nonumber \\
& + \frac{\left[a^{'}(t)\right]^{2} a(t)^{-\frac{1}{m+1}-2}}{2(2m+2)^{2}}\left[g\left(\int^{t}_{t_0} a(x)^{\frac{1}{m}}dx\right)\right]^{2} \nonumber \\
& - \frac{a^{'}(t) a(t)^{-1}}{2m+2} g\left(\int^{t}_{t_0} a(x)^{\frac{1}{m+1}}dx\right) g^{'}\left(\int^{t}_{t_0} a(x)^{\frac{1}{m+1}}dx\right).
\eea
The function $g$ and its derivative $g^{'}=dg/dt$ are periodic and bounded functions, 
being the solutions of the time-independent Hamiltonian system (\ref{ode}). 
Assuming that $a(t)$ is an unbounded monotonically increasing
function of time, and also that  the $\frac{d\log(a(t))}{dt}$ 
is a decreasing function of time, 
we see that if $t\rightarrow\infty$, then $(q(t),p(t))\rightarrow(0,\infty)$.
The energy depends on the initial 
conditions $\phi_0$ and $E_0$, through the expression in the rectangular brackets,
and this is the term which remains non-vanishing for large time $t\rightarrow\infty$,
\bea
E(t) \simeq c(\phi_0, E_0) \;a(t)^{\frac{1}{m+1}}.
\eea
After calculating the mean energy and the mean square of the energy, and dropping 
the decaying terms, we arrive at the following
expressions of the leading terms for the average energy and
the variance, 
$Var(E(t))=\langle E(t)^{2} \rangle - \langle E(t) \rangle^{2}$, 
in the limit $t\rightarrow\infty$,
\bea \label{av}
\langle E(t) \rangle\simeq \frac{ a(t)^{\frac{1}{m+1}}}{2\pi} \int_{0}^{2\pi} c(\phi_0, E_0) d\phi_0 ,
\eea
\bea \label{var}
Var(E(t)) \simeq \frac{a(t)^{\frac{2}{m+1}}}{2\pi} \left[ \int_{0}^{2\pi} (c(\phi_0, E_0))^2 d\phi_0 -\frac{1}{2\pi}\left( \int_{0}^{2\pi} c(\phi_0, E_0) d\phi_0 \right)^2 \right].
\eea
Using the previous expression for the average energy we  
calculate also the action ratio \cite{ABR} of the final and initial action. 
The general formula for the action at the mean energy is,
\bea  \label{action}
I(\langle E(t) \rangle) = \frac{1}{2\pi} \oint_{H(q,p)=<E(t)>} p\;dq = \frac{2^{\frac{m+1}{2 m}} m^{\frac{1}{2 m}}\Gamma\left(\frac{1}{2 m}\right)}{2 m \sqrt{\pi}(a(t))^{\frac{1}{2 m}}\Gamma\left(\frac{3 m + 1}{2 m}\right)}\langle E(t)\rangle^{\frac{m+1}{2 m}}.
\eea
hence the action ratio for large times $t\rightarrow\infty$ is,
\bea \label{AcRa}
\frac{I(\langle E_1 \rangle)}{I(E_0)} \simeq (a(t_0))^{\frac{1}{2m}} \left[\frac{\int_{0}^{2\pi} c(\phi_0, E_0) d\phi_0}{2\pi E_0}\right]^{\frac{m+1}{2 m}}=const.
\eea
This is an important result, showing that the action 
ratio remains constant for large times. The numerical empirics shows that
it is always slightly larger than its initial value.
 
\section{Examples} \label{ex}

In this section we show two examples, the linear oscillator and the quartic 
oscillator. In the linear oscillator we have exact analytical results for 
the energy distribution function, as well as for the average energy, 
the variance and the action ratio 
\cite{RR2006a,RR2006b,RRS2006,KR2007,RR2008}, but reproduce them here
using the more general, although only  leading-order,
 nonlinear WKB-like method \cite{PapRob2012}, 
now for arbitrarily large times.
In the quartic oscillator it is not easy to have explicit 
theoretical results due the complexity of the Jacobi elliptic functions,
but using the numerical technique, for the
case with $a(t)= 1 + \epsilon t$, we calculate and analyze the solutions 
and show that the general theory of section \ref{General} is correct.
For the higher order potentials, $m=3,4,\dots$, we calculate numerically
the energy distribution as a function of time (originating from a
microcanonical ensemble of initial conditions)  and its mean value
and the variance, and verify whether they obey the theoretical power
laws from section \ref{General}. 

\subsection{Linear oscillator} \label{LO}

We consider the Hamiltonian of the linear oscillator
in the following form,
\bea \label{LH}
H(q,p,t) = \frac{p^{2}}{2} + \omega^{2}(t)\frac{q^{2}}{2},
\eea
where, $\omega(t)$ is an unbounded monotonically increasing function of time.
The general solution using the general nonlinear WKB-like method
(which of course is identical to the linear WKB method in the leading order) to the 
first order approximation is,
\bea
q(t)&=\frac{c_{1}}{\sqrt{\omega(t)}} \sin(c_{2} + \int^{t}_{t_0}\omega(x) dx), \nonumber \\
p(t)&=c_{1} \sqrt{\omega(t)} \cos(c_{2} + \int^{t}_{t_0}\omega(x) dx) - \frac{c_1 \omega^{'}(t)}{2(\omega(t))^{\frac{3}{2}}} \sin(c_{2} + \int^{t}_{t_0}\omega(x) dx),
\eea
where, $c_{1}$ and $c_{2}$ are integration constants.
The action-angle variables, for $t=t_{0}$, are
\bea
q_{0} = \frac{\sqrt{2E_{0}}}{\omega_{0}}\; \sin\phi_{0}, \quad \quad p_{0} = \sqrt{2E_{0}}\;\cos\phi_{0},
\eea
where, $\omega_0 = \omega(t_0)$, $E_0$ is the initial energy 
and $\phi_0$ is the initial angle. Using this we get,
\bea
q(t)&=\frac{b_{1}}{\sqrt{\omega(t)}}\left[(b_{2} + \cot\phi_{0})\sin\phi_{0}\;\sin\left(\int^{t}_{t_{0}} \omega(x)dx \right) + \sin\phi_{0}\;\cos\left(\int^{t}_{t_{0}} \omega(x)dx \right)\right], \nonumber \\
p(t)&=b_{1}\sqrt{\omega(t)}\left[(b_{2} + \cot\phi_{0})\sin\phi_{0}\;\cos\left(\int^{t}_{t_{0}} \omega(x)dx \right) - \sin\phi_{0}\;\sin\left(\int^{t}_{t_{0}} \omega(x)dx \right)\right] \nonumber \\
& - \frac{b_{1}\omega^{'}(t)}{2(\omega(t))^{\frac{3}{2}}}\left[(b_{2} + \cot\phi_{0})\sin\phi_{0}\;\sin\left(\int^{t}_{t_{0}} \omega(x)dx \right) + \sin\phi_{0}\;\cos\left(\int^{t}_{t_{0}} \omega(x)dx \right)\right]. \nonumber \\
& 
\eea
where, $b_{1}=\sqrt{\frac{2E_{0}}{\omega_{0}}}$ and $b_{2}=\frac{\omega^{'}_{0}}{2\omega_{0}^{2}}$.
The average energy after the integration over the $\phi_0$ in 
the interval $\left[0, 2\pi\right]$ is,
\bea  \label{linaveenergy}
\langle E(t) \rangle &= \left(\frac{b_{1}^{2}}{2} + \frac{1}{4} b_{1}^{2} b_{2}^{2}\right) \omega(t) -\frac{b_{1}^{2} b_{2}^{2}(\omega^{'}(t))^{2}\cos\left(2\int^{t}_{t_{0}} \omega(x)dx \right)}{32\omega^{3}(t)} + \frac{b_{1}^{2} b_{2}^{2}(\omega^{'}(t))^{2}}{32\omega^{3}(t)} \nonumber \\
&- \frac{b_{1}^{2} b_{2}^{2}\omega^{'}(t)\sin\left(2\int^{t}_{t_{0}} \omega(x)dx \right)}{8\omega(t)}  + \frac{b_{1}^{2} b_{2}(\omega^{'}(t))^{2}\sin\left(2\int^{t}_{t_{0}} \omega(x)dx \right)}{16\omega^{3}(t)} \nonumber \\
&- \frac{b_{1}^{2} b_{2}\omega^{'}(t)\cos\left(2\int^{t}_{t_{0}} \omega(x)dx \right)}{4\omega(t)} + \frac{b_{1}^{2}(\omega^{'}(t))^{2}}{16\omega^{3}(t)}.
\eea
For large time and after a long calculation of the mean squared energy, 
and dropping the constant and decaying terms, 
we arrive at the following expressions,
\bea
&\langle E(t) \rangle = \left(\frac{b_{1}^{2}}{2} + \frac{1}{4} b_{1}^{2} b_{2}^{2}\right) \omega(t), \nonumber \\
&Var(E(t)) = \frac{1}{32} b_{1}^{4} b_{2}^{2} (4 + b_{2}^{2}) \omega^{2}(t).
\eea
While the higher order terms of $\langle E(t) \rangle$ are 
clearly visible in equation (\ref{linaveenergy}), for the
variance they are too complex to be presented here, but we only
mention that they as a function of time through $\omega (t)$  
are constant in the next order, and  decay as $1/\omega^2(t)$ in the
next next order. In manipulations
it is necessary to use the computer symbolic calculations.

\subsection{Quartic oscillator} \label{QO}

We consider the Hamiltonian of the quartic oscillator 
in the following form,
\bea \label{QH}
H(q,p,t) = \frac{p^{2}}{2} + a(\lambda)\frac{q^{4}}{4},
\eea
where $a(t)$ is an unbounded monotonically increasing function of time.
The general solution using the nonlinear WKB-like method \cite{PapRob2012} to the 
first order approximation is,
\bea
q(t)&= \frac{\sqrt{2}\;c_{1}}{(a(t))^{\frac{1}{6}}}\;{\rm sn}\left(c_{2} + c_{1}\int^{t}_{t_0}(a(x))^{\frac{1}{3}} dx,i\right), \nonumber \\
p(t)&= \sqrt{2}\;c_{1}^{2}\;(a(t))^{\frac{1}{6}}\sqrt{1-{\rm sn}^{4}\left(c_{2} + c_{1}\int^{t}_{t_0}a(x)^{\frac{1}{3}} dx,i\right)} \nonumber \\
&-\frac{\sqrt{2}\;c_{1}\;a^{'}(t)\; {\rm sn}\left( c_{2} + c_{1}\int^{t}_{t_0}a(x)^{\frac{1}{3}} dx,i\right)}{6\; a(t)^{\frac{7}{6}}},
\eea
where, $c_{1}$ and $c_{2}$ are integration constants. 
By $sn$ we denote the Jacobi elliptic function \cite{Olver}.
The initial conditions in terms of the action-angle  variables, 
for $t = t_{0}$, are
\bea
   q_{0} = \sqrt[4]{\frac{4E_{0}}{a_{0}}}\; {\rm sn}\left( \frac{\Gamma^{2}\left(
   \frac{1}{4}\right)}{2^{\frac{3}{2}}\pi^{\frac{3}{2}}}\phi_0,i\right), \quad p_
   {0} = \sqrt{2E_{0}}\sqrt{1-{\rm sn}^{4}\left( \frac{\Gamma^{2}\left( \frac{1}
   {4}\right)}{2^{\frac{3}{2}}\pi^{\frac{3}{2}}}\phi_0,i\right)},
\eea
where $a_0 = a(t_0)$, $E_0$ is the initial energy 
and $\phi_0$ is the initial angle.
We calculate the constants $c_{1}, c_{2}$ as functions of 
the initial conditions as follows,
\bea
c_2 = {\rm sn}^{-1}\left( \frac{q_{0}\; a_{0}^{\frac{1}{6}}}{\sqrt{2}\;c_1}, i\right), \nonumber \\
c_1 = \sqrt[4]{\frac{E_{0}}{a_{0}^{\frac{1}{3}}} + \frac{\sqrt{2}q_{0}\;p_{0}\;a_{0}^{'}}{6\;a_{0}^{\frac{4}{3}}} + \frac{q_{0}^{2}\;(a_{0}^{'})^{2}}{72\; a_{0}^{\frac{7}{3}}}},
\eea
and obtain the following expression for the energy
\bea
E(t) = c_{1}^{4}(a(t))^{\frac{1}{3}} + \frac{c_{1}^{2}(a^{'}(t))^{2} A^{2}(t)}{36\; (a(t))^{\frac{7}{3}}} - \frac{c_{1}^{3}a^{'}(t)A(t)\sqrt{1 - A^{4}(t)}}{3\;a(t)},
\eea
where the function $A(t)$ is the following periodic function,
\bea
 A(t)={\rm sn}\left(c_{2} + c_{1}\int^{t}_{t_0}(a(x))^{\frac{1}{3}} dx,i\right).
\eea
We cannot analytically calculate the average energy 
from this expression, unlike in the linear oscillator case.
But we see that the leading order term, for $t\gg 1$, 
is proportional to the power $(a(t))^{\frac{1}{3}}$. The
$E^{2}(t)$ has leading order, for $t\gg1$, proportional to
$(a(t))^{\frac{2}{3}}$. For the average energy and
the variance, dropping the decaying terms 
in the limit $t\rightarrow\infty$, we find
\bea \label{quaaveenergy}
&\langle E(t) \rangle = \frac{a(t)^{\frac{1}{3}}}{2 \pi}\int_{0}^{2\pi} c_{1}^{4} d\phi_{0}, \nonumber \\
&Var(E(t))=\frac{a(t)^{\frac{2}{3}}}{2\pi}\left[\int_{0}^{2\pi} c_{1}^{8} d\phi_{0}-\frac{1}{2\pi}\left(\int_{0}^{2\pi} c_{1}^{4} d\phi_{0}\right)^{2}\right].
\eea
The variance is nonzero due to the fact that 
$\left(\int_{0}^{2\pi} c_{1}^{4} d\phi_{0}\right)^{2} < 2\pi\int_{0}^{2\pi} c_{1}^{8} d\phi_{0}$. The higher order constant and decaying terms 
in equations (\ref{quaaveenergy}) can be easily calculated, 
but are too complex to be displayed here. In manipulations
it is necessary to use the computer symbolic calculations.

\begin{figure} 
\center 
    \includegraphics[width=15cm]{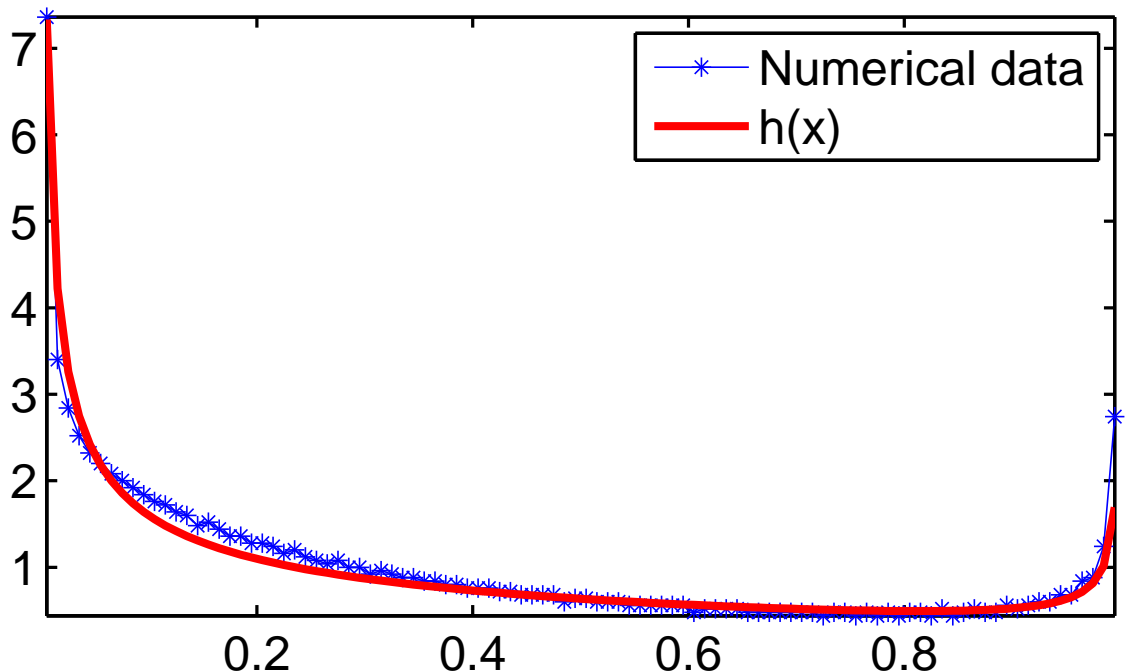} 
    \caption{The fitting probability density function $h(x)$ for the 
quartic oscillator, for $a(t)=1 + \epsilon t$, for $\epsilon = 1$, $t=2000$ 
and initial energy $E_0 = 0.5$.}
\label{actions4}
\end{figure}

The energy distribution for the general family of systems (arbitrary $m$) is 
difficult to find analytically in a closed formula, 
except for the linear oscillator $m=1$ \cite{RRS2006}.
For the model $a(t) =1 +\epsilon t$, we have numerically calculated the energy
distribution functions and found that for all of them (any $m$) it reminds of 
the arcsine distribution, as it has two integrable singularity spikes at the minimal 
and maximal energy,
but is distorted in an asymmetric manner. For the quartic
oscillator we can find an empirical approximation. Namely, we propose the 
following rough approximation for the probability density, which is
linearly distorted arcsine distribution
\bea
h(x) = (\frac{1}{\pi} -\frac{\alpha}{2} + 
\alpha x) x^{-\frac{1}{2}}(1-x)^{-\frac{1}{2}}, \quad x\in[0,1].
\eea
In the specific numerical example shown in figure \ref{actions4}, we empirically 
(by best fitting) find $\alpha = 0.40175$. 
We remark that $\alpha = 0$ yields the arcsine distribution, 
which is exact for the linear oscillator for any time dependence $\omega(t)$.

\subsection{Higher power law homogeneous potentials} \label{HHP}

In general it is not easy to calculate the exact expressions of the mean 
energy and the variance for higher order homogeneous power law potentials, 
because of the complicated dependence on the initial conditions, due to the
complexity of the fundamental solution $g(t)$, entering into the WKB formulae, 
as described in section \ref{General}. Nevertheless, the power laws for the 
mean energy (\ref{av})
and the variance (\ref{var}) are predicted explicitly by the theory, and
the proportionality
constants can be found empirically (numerically), if needed.
We again use the specific model $a(t)= 1 + \epsilon t$, with $\epsilon=1$,
for $m=1,\ldots,6$,
and perform the numerical integrations for very large time $t=2000$,
using the 8th order symplectic integrator.
In the table 1 we present the comparison between the theoretical exponents
$1/(m+1)$ for the mean energy, and in the table 2  the exponents
$2/(m+1)$ for the variance, with their corresponding numerical values. 
In addition, in table 3 we calculate the action ratio, using equations
\eref{av} and \eref{AcRa}.

\begin{table}
\center
    \begin{tabular}{ | l | l | l | p{5cm} |}
    \hline
    \multicolumn{4}{|c|}{The exponents for $\langle E(t) \rangle$} \\
    \hline
    m & Numerical value & Theoretical value & Error\\ \hline
    1 & 0,4996481 & 0.5 & 3$\cdot10^{-4}$\\ \hline
    2 & 0.3330987 & 0.3333333 & 3$\cdot10^{-4}$\\ \hline
    3 & 0.249824 & 0.25 & $10^{-4}$\\ \hline
    4 & 0.1998592 & 0.2 & $10^{-4}$\\ \hline
    5 & 0.1665493 & 0.1666666 & $10^{-4}$\\ \hline
    6 & 0.1427567 & 0.1428571 & $10^{-4}$ \\ \hline  
    \end{tabular}\\ 
    \caption{We show the power law exponents for $\langle E(t) \rangle$, 
    whose theoretical value is $1/(m+1)$, as derived in ~\eref{av}, 
    compared to the numerical results; initial energy $E_0 = 0.5$. 
    The number of points in the initial ensemble is 5000. Time is $t=2000$.}
\end{table}
\begin{table}
\center
    \begin{tabular}{ | l | l | l | p{5cm} |}
    \hline
    \multicolumn{4}{|c|}{The exponents for $Var(E(t))$} \\
    \hline
    m & Numerical value & Theoretical value & Error\\ \hline
    1 & 0,9992962 & 1 & 7$\cdot10^{-4}$\\ \hline
    2 & 0,6661974 & 0.6666666 & 4$\cdot10^{-4}$\\ \hline
    3 & 0,499648 & 0.5 & 3$\cdot10^{-4}$\\ \hline
    4 & 0,3997184 & 0.4 & 2$\cdot10^{-4}$\\ \hline
    5 & 0,3330986 & 0.3333332 & 2$\cdot10^{-4}$\\ \hline
    6 & 0.2855071 & 0.2857142 & 2$\cdot10^{-4}$\\ \hline
    \end{tabular}\\
    \caption{We show the power law exponents for $Var(E(t))$, 
    whose theoretical value is $2/(m+1)$, as derived in ~\eref{var}, 
    compared to the numerical results; initial energy $E_0 = 0.5$. 
    The number of points in the initial ensemble is 5000. Time is $t=2000$.}
\end{table}
\begin{table}
\center
    \begin{tabular}{ | l | l | l | p{5cm} |}
    \hline
    \multicolumn{4}{|c|}{Values for the action ratio} \\
    \hline
    m & Numerical value & Theoretical value & Error\\ \hline
    1 & 1.0218739 & 1.0190229 & 2$\cdot10^{-3}$\\ \hline
    2 & 1.0110728 & 1.009601 & $10^{-3}$\\ \hline
    3 & 1.0061271 & 1.0051507 & 9$\cdot10^{-4}$\\ \hline
    4 & 1.003837 & 1.0031063 & 7$\cdot10^{-4}$\\ \hline
    5 & 1.0026292 & 1.002045 & 6$\cdot10^{-4}$\\ \hline
    6 & 1.0019214 & 1.0014258 & 5$\cdot10^{-4}$ \\ \hline
    \end{tabular}\\
   \caption{We show the exact value of the action ratio compared to the theoretical, 
   as derived in ~\eref{AcRa} for the initial energy $E_0 = 0.5$. 
   The number of points in the initial ensemble is 5000. Time is $t=2000$.}
\end{table}

\section{Discussion and conclusion}

In this work we have analyzed the statistical properties of the one degree
of freedom  time-dependent Hamilton systems with homogeneous power law potential.
We have calculated numerically the energy distribution as a function of time,
for large times. 
Our main interest is in the value of the final average energy, 
the variance and the action at the final average energy.
These are the main parameters describing the energy layer which
evolves from the initial microcanonical ensemble.
In the case that the dependence on time is monotonic and unbounded, 
we have calculated the average energy and the variance 
of the energy using the nonlinear WKB method \cite{PapRob2012},
and derived the power laws observed in the asymptotic limit of large times.
In particular, for the adiabatic invariant (action) we have proven that
the value becomes constant for very large time but is (empirically)
larger than the initial value.
The agreement between the theory and the numerics is excellent. 
Unlike this limit, for small or intermediate times, 
the analytic expressions become too complicated to be expressed
in a closed formula. 

To the best of our knowledge this is the first explicit
application of the nonlinear WKB-like theory developed in
\cite{PapRob2012}. There are of course many important open questions, 
namely how to describe other 1D nonlinear Hamilton oscillators,
like those studied e.g. in \cite{ABR}, 
possibly for arbitrary drivings $a(t)$, for which the nonlinear
WKB-like method should be generalized, and possibly improved 
beyond the leading order.

\section*{Acknowledgements}

Financial support of the Slovenian Research Agency ARRS
under the grant P1-0306 is gratefully acknowledged.

\end{document}